\documentclass[11pt]{article}

\usepackage{amsmath}
\usepackage{amsthm}
\usepackage{amssymb}
\usepackage{amscd}
\usepackage{color}

\theoremstyle{definition}
\newtheorem{theorem}{Theorem}[section]
\newtheorem{prop}[theorem]{Proposition}
\newtheorem{lemma}[theorem]{Lemma}
\newtheorem{corollary}[theorem]{Corollary}

\newtheorem{example}[theorem]{Example}
\newtheorem{remark}[theorem]{Remark}

\numberwithin{equation}{section}
\newcommand\bea{\begin{eqnarray}}
\newcommand\ena{\end{eqnarray}}
\newcommand\non{\nonumber}

\allowdisplaybreaks

\title
{Tau Functions of $(n,1)$ curves and Soliton Solutions 
on Non-Zero Constant Backgrounds
  \\
}

\author{
Atsushi Nakayashiki
\footnote{
Department of Mathematics, Tsuda University,
Kodaira, Tokyo 187-8577, Japan,
{\tt atsushi@tsuda.ac.jp}
}
}

\date{
}

\begin{document}

\maketitle

\begin{abstract}
We study the tau function of the KP-hierarchy associated with an $(n,1)$ curve $y^n=x-\alpha$. 
If $\alpha=0$ the corresponding tau function is $1$. On the other hand if $\alpha\neq 0$ 
the tau function becomes the exponential of a quadratic function of the time variables. 
 By applying vertex opertaors to the
latter we obtain soliton solutions on non-zero constant backgrounds.

\end{abstract}

\section{Introduction}
In our previous paper \cite{N2020} the degeneration of the theta function solution (tau function)
of the KP-hierarchy as a result of  what we call  one step degeneration of an $(n,s)$ curve 
for $n=2,3$ has been studied. The obtained formula expresses the degenerate tau function as a sum of 
combinations of exponential functions with tau functions of a lower genus curve.
Repeating this process we finally come to the tau function associated with an
$(n,1)$ curve of genus zero. In this paper we  derive the explicit formula for the tau function 
of an $(n,1)$ curve with an arbitrary $n\geq 2$. We use it as a seed solution in the vertex operator construction of solutions of the KP-hierarchy.
 In the case of $n=2$ the solutions obtained in this way are considered as 
 solitons on non-zero constant backgrounds.

For $n\geq 2$  consider the rational algebraic curve $C_n$ defined by
\bea
&&
y^n=x-\alpha,
\label{n1curve-intro}
\ena
which we call an $(n,1)$ curve.
From this curve a solution of the KP-hierarchy (tau function) is constructed as follows.
To this end we use the Sato Grassmannian which we denote by UGM (=universal Grassmann manifold)
after Sato \cite{SS,SN}. The  Sato Grassmannian is the parameter space of solutions of the KP-hierarchy and is
defined as the set of certain subspaces of the vector space $V={\mathbb C}((z))$ of Laurent series in a variable $z$.

Consider the point $\infty$ of $C_n$ and take the local coordinate $z$ around 
$\infty$ such that
\bea
&&
x=z^{-n},
\hskip5mm
y=z^{-1}(1+O(z)).
\non
\ena
This choice of the local coordinate is necessary in order to study the degeneration of solutions of the 
KP-hierarchy associated with an $(n,s)$ curve \cite{N2020}.
Define the vector space $V_n$ by
\bea
&&
V_n=\sum_{j=0}^\infty {\mathbb C} y^j.
\non
\ena
By expanding $y^j$ in the coordinate $z$ $V_n$ can be considered as a subspace of $V$.
 Then it is known that it belongs to UGM (see \cite{N2019} for example). 
 We denote this point of UGM by $V_n(z)$ meaning that it is obtained from $V_n$ using the coordinate $z$.

For each point of UGM a solution $\tau({\bf t})$ of the KP-hierarchy is constructed, up to constant,
 in the form of the Schur function expansion as
\bea
&&
\tau({\bf t})=\sum_{\lambda}\xi_\lambda s_\lambda({\bf t}),
\non
\ena
where the summation is over all partitions, $s_\lambda({\bf t})$ is the Schur function corresponding to $\lambda$ and
$\xi_\lambda$ the Pl\"ucker coordinate of the point of UGM.

In the present case the Schur function expansion of the tau function corresponding to $V_n(z)$ begins 
from a non-zero constant. We define the tau function $\tau_0({\bf t};V_n(z))$ as that corresponding to 
$V_n(z)$ normalized as 
\bea
&&
\tau_0({\bf t};V_n(z))=1+\sum_{|\lambda|>0}\xi_\lambda s_\lambda(t).
\label{tau0-intro}
\ena

In the case $\alpha=0$ we easily have $\tau_0({\bf t};V_n(z))=1$. So the actual problem here is to compute
$\tau_0({\bf t};V_n(z))$ for $\alpha\neq 0$. 
The result is 
\bea
&&
\tau_0({\bf t};V_n(z))=e^{\frac{1}{2}\left(q({\bf t})+L({\bf t})\right)},
\label{main-intro}
\ena
where $q({\bf t})$ is a quadratic form in ${\bf t}={}^t(t_1,t_2,t_3,...)$ and $L({\bf t})$ is a linear 
form in ${\bf t}$ (see Theorem \ref{main}).
It is difficult to prove such a formula directly from the Schur function expansion. 
The strategy here to derived the formula (\ref{main-intro}) is to consider a coordinate change 
of the corresponding wave and the adjoint wave function \cite{DJKM}.
This idea comes from the paper \cite{DN} by Dubrovin and Natanzon,
who first studied the effect of a coordinate change in the construction of 
 solutions of the KP-equation.

The Schur function expansion of the tau function (\ref{main-intro}) is also calculated explicitly 
by normalizing a frame of $V_n(z)$ (see Theorem \ref{main2}). 
We use the Giambelli formula for Pl\"ucker coordinates \cite{EH,NOS} in this computation.

The second logarithmic derivative of a tau function with respect to $t_1$ (see (\ref{KP-2})) gives a 
solution of the KP-equation (\ref{KP-3}). The solution corresponding to $\tau_{0}(t;V_n(z))$ is 
a constant. So (\ref{main-intro}) itself does not look an interesting solution of the KP equation, although 
it is used to describe the degeneration of theta function solution as mentioned in the beginning.
However this is not the case.  It plays a role as a seed to create various solutions of the KP-hierarchy.
In fact it is well known that soliton solutions of the KP-hierarchy are constructed by applying vertex operators 
to the trivial solution $1$ \cite{DJKM}. 
Similarly we also get various solutions by applying vertex operators to $\tau_{0}(t;V_n(z))$.
In particular, if  $n=2$ and $\alpha\neq0$ solutions obtained in this way are considered as solitons 
with non-zero constant asymptotics (see Corollary \ref{vop-cor} and Theorem \ref{vop-main}).
Note that for this computation to work  it is indispensable to introduce 
the infinite number of time variables.

Finally we mention that in the paper \cite{Givental2003} the tau function of the form (\ref{main-intro}) had been studied in relation with Gromov-Witten invariants and the dispertionless KP-hierarchy. In that paper 
all points of UGM corresponding to tau functions of the n-reduced KP-hierarchy, which are expressed in the 
form of the exponential of a quadratic form,  are determined. The point $V_n(z)$ is a special family, depending on the parameter $\alpha$, in them. However such explicit formula as in Theorem \ref{main} of the tau function was not derived in \cite{Givental2003} as far as the author understands.

The paper is organized as follows. After the introduction a brief review on the Sato Grassmannian and the KP-hierarchy are given in section 2. In section 3 the problems and main results are stated. Sections 4 to 7 are  devoted to the proofs of main theorems. In section 4 the problem of determining the tau function 
is reformulated in terms of the wave function. Beginning from the trivial wave function corresponding to the 
trivial tau function $1$, the condition for the tau function which we seek for is formulated by the 
coordinated changed the wave and the adjoint wave function. 
The series expansion of the function which appears 
in the wave function in question is determined in section 5. Based on the results of section 4 and 5 Theorem 
\ref{main} is proved in section 6. 
In section 7 Theorem \ref{main2} is proved  by  computing the Pl\"ucker coordinates of $V_n(z)$.
The generating function of the coefficients of 
the quadratic form $q(t)$ is computed in section 8. Here the genus zero analogue of the bilinear meromorphic differential 
\cite{BEL1997, BEL1999, N2010-1} of an $(n,s)$ curve with positive genus plays a crucial role.
In section 9 soliton solutions on non-zero constant backgrounds 
are computed  by applying vertex operators to $\tau_0({\bf t};V_n(z))$. 
In this calculation the result of section 8 is crucial.

\section{Sato Grassmannian}
In this section we beriefly review the theory of the Sato Grassmannian and the KP-hierarchy.

Let $\tau({\bf t})$ be a function of  ${\bf t}={}^t(t_1,t_2,...)$.
The KP-hierarchy is the  equation for $\tau({\bf t})$ given by 
\bea
&&
{\rm Res}_{k=\infty} \tau({\bf t}-{\bf s}-[k^{-1}])({\bf t}+{\bf s}+[k^{-1}])e^{-2\sum_{j=1}^\infty s_j k^j}\frac{dk}{2\pi i}=0,
\label{KP-1}
\ena
where ${\bf s}={}^t(s_1,s_2,...)$, $[k]={}^t(k,k^2/2,k^3/3,...)$. 
By expanding in $s_j$, $j\geq 1$, (\ref{KP-1}) gives an 
infinite number of differential equations for $\tau({\bf t})$ expressed in Hirota's bilinear form \cite{DJKM}.
Set $(x,y,t)=(t_1,t_2,t_3)$ and, 
for a solution $\tau({\bf t})$ of (\ref{KP-1}) , 
\bea
&&
u({\bf t})=2\frac{\partial^2}{\partial x^2}\log \tau({\bf t}).
\label{KP-2}
\ena
Then $u({\bf t})$ is a solution of the KP-equation,
\bea
&&
3u_{yy}+(-4u_{t}+6uu_{x}+u_{xxx})_{x}=0.
\label{KP-3}
\ena
The totality of solutions of the KP-hierarchy constitues a certain infinite dimensional Grassmann manifold
called the Sato Grassmannian \cite{SS}. Let us recall its definition and fundamental properties.

Let $V={\mathbb C}((z))$ be the vector space of Laurent series in $z$. Define two subspaces of $V$ by 
$V_\phi={\mathbb C}[z^{-1}]$, $V_0=z{\mathbb C}[[z]]$.
Then $V=V_\phi\oplus V_0$. Let $\pi:V\longrightarrow V_\phi$ be the projection
map. Then the Sato Grassmannian which we denote by UGM (=universal Grassmann manifold) after Sato \cite{SS,SN}, is defined as the set of  subspaces $U$ of  $V$ such that 
\bea
&&
\dim( {\rm Ker}\, \pi|_U)=\dim ({\rm Coker}\, \pi|_U)<\infty.
\non
\ena

\begin{example} The subspace $V_\phi$ belongs to UGM. In this case 
\bea
&&
\dim( {\rm Ker}\, \pi|_{V_\phi})=\dim ({\rm Coker}\, \pi|_{V_\phi})=0.
\label{dim=0}
\ena.
\end{example}

In general a subspace $U$ defined by
\bea
&&
U=\sum_{j\leq 0}{\mathbb C}\xi_j,
\quad
\xi_{j}=z^{j}+\sum_{i=1}^\infty \xi_{i,j} z^i,
\quad
\xi_{i,j}\in{\mathbb C},
\label{big-cell-1}
\ena
satisfies (\ref{dim=0}) and belongs to UGM. 
The totality of points of UGM corresponding to such frames is denoted by ${\rm UGM}^\phi$, which 
forms a cell called the big cell of UGM.

A point $U$ of UGM can be specified by its frame, that is,  a basis of $U$.
If  we associate the infinite column vector $(a_n)_{n\in {\mathbb Z}}$ to an element $\sum a_n z^n$ of $V$,
a frame of $U$ can be expressed by an ${\mathbb Z}\times {\mathbb Z}_{\leq0}$ matrix 
$\xi=(\xi_{i,j})_{i\in {\mathbb Z},j \in {\mathbb Z}_{\leq0}}$, where ${\mathbb Z}_{\leq 0}$ denotes 
the set of non-positive integers.

In writing the matrix $\xi$ we follow the usual convention that the row numbers increase downward and 
the column numbers increase rightward.
For example the frame  $\xi=(...,\xi_{-1},\xi_0)$ of $U$ given by  (\ref{big-cell-1})  is represented as 
\bea
&&
\xi=
\left(
\begin{array}{ccc}
\quad&\vdots&\vdots\\
\cdots&1& 0 \\
\cdots& 0 & 1 \\
---&---&---\\
\cdots&\xi_{1,-1}&\xi_{1,0}\\
\cdots&\xi_{2,-1}&\xi_{2,0}\\
\quad&\vdots&\vdots\\
\end{array}
\right).
\label{big-cell}
\ena

In general it is always possible to take a frame satisfying the following condition;
there exists a negative integer $l$ such that
\bea
&&
\xi_{i,j}=\left\{
\begin{array}{cl}
1&\text{ if $j<l$ and $i=j$ }\\
0&\text{ if ($j<l$ and $i<j$) or ($j\geq l$ and $i<l$)}.
\end{array}
\right.
\label{frame-cond}
\ena
In the sequel we always take a frame which satisfies this condition, although it is not unique.

A Maya diagram $M=(m_j)_{j=0}^\infty$ is a sequence of intergers  such that $m_0>m_1>\cdots$ and $m_j=-j$ for all sufficiently large $j$. For a Maya diagram $M=(m_j)_{j=0}^\infty$ the corresponding partition is defined by 
$\lambda(M)=(j+m_j)_{j=0}^\infty$. By this correspondence the set of Maya diagrams and the set 
of partitions bijectively correspond to each other. 
We identify a partition with the corresponding Maya diagram.

For a frame $\xi$ and a Maya diagram $M=(m_j)_{j=0}^\infty$ define the Pl\"ucker coordinate of $\xi$ corresponding to $M$ by
\bea
&&
\xi_M=\det(\xi_{m_i,j})_{-i,j\leq0}.
\non
\ena
If $M$ corresponds to a partition $\lambda$, $\xi_M$ is denoted also by $\xi_\lambda$.
Due to the condition (\ref{frame-cond}) and the condition of the Maya diagram
 this infinite determinant can be computed by the finite determinant $\det(\xi_{m_i,j})_{k\leq -i,j\leq0}$ for
 a sufficiently small $k$.

Define the  Schur function $s_{(n)}(t)$ corresponding to the partition $(n)$ by
\bea
&&
e^{\sum_{n=1}^\infty t_n\kappa^n}=\sum_{n=0}^\infty s_{(n)}(t) \kappa^n.
\non
\ena
and the Schur function \cite{Mac1995} corresponding to a partition $\lambda=(\lambda_1,...,\lambda_l)$  by
\bea
&&
s_\lambda({\bf t})=\det(s_{(\lambda_i-i+j)}(t))_{1\leq i,j\leq l}.
\non
\ena
We assign weight $j$ to the variable $t_j$. Then $s_\lambda({\bf t})$ becomes homogeneous 
of  weight $|\lambda|=\lambda_1+\cdots+\lambda_l$.
 
To a point $U$ of UGM take a frame $\xi$ of $U$ and define the tau function by
\bea
&&
\tau({\bf t};\xi)=\sum_\lambda \xi_\lambda s_{\lambda}({\bf t}),
\label{tau}
\ena
where the sum is taken over all partitions.
We call it the Schur function expansion of $\tau({\bf t};\xi)$.

If we change the frame $\xi$, Pl\"ucker coordinates and consequently 
 the tau function are multiplied by a non-zero constant due to the property of a determinant.
We call $\tau({\bf t};\xi)$, for any frame $\xi$ of $U$, a tau function of $U$.
So tau functions of a point of UGM differ by non-zero constant multiples to each other.

For a frame $\xi$ of the form (\ref{big-cell}), (\ref{big-cell-1}) $\xi_{(0)}=1$ and the Schur function 
expansion takes the form
\bea
&&
\tau({\bf t};\xi)=1+\text{p.w.t},
\label{big-cell-2}
\ena
where p.w.t. means positive weight terms. For $U\in {\rm UGM}^\phi$ we denote $\tau_0({\bf t};U)$ 
the tau function normalized as in (\ref{big-cell-2}). We call it the normalized tau function of $U$.

The fundamental theorem of the Sato theory is the following \cite{SS}
 (see also \cite{SN, KNTY, Mul1994}).

\begin{theorem}\cite{SS}\label{Sato-Sato}
For a frame $\xi$ of a point of UGM $\tau({\bf t};\xi)$ is a solution of the KP-hierarchy.
Conversely for a 
formal power series solution $\tau({\bf t})$ of the KP-hierarchy there exists a unique point $U$
of UGM such that $\tau({\bf t})$ is a tau function of $U$.
\end{theorem}

The inverse construction from a solution $\tau({\bf t})$ of the KP-hierarchy  to the point $U$ of UGM is given using the wave function as follows \cite{SS,SN,KNTY,N2010-2}.
The wave function $\Psi^\ast({\bf t};z)$ and the adjoint wave function  $\Psi^\ast({\bf t};z)$ \cite{DJKM} corresponding to $\tau({\bf t})$ are defined by 
\bea
&&
\Psi({\bf t};z)=\frac{\tau({\bf t}-[z])}{\tau({\bf t})}e^{\sum_{i=1}^\infty t_iz^{-i}}.
\quad
\Psi^\ast({\bf t};z)=\frac{\tau({\bf t}+[z])}{\tau({\bf t})}e^{-\sum_{i=1}^\infty t_iz^{-i}}.
\label{wave}
\ena
These functions are solutions of the linear problem associated with the KP-hierarchy \cite{DJKM}.
For the inverse construction we use the adjoint wave function. 
Let $\Psi_i^\ast(z)$ be the Laurent series in $z$ defined by
\bea
\left(\tau({\bf t})\Psi^\ast({\bf t};z)\right)\vert_{t=(x,0,0,0,...)}
=
\sum_{i=0}^\infty \Psi_i^\ast(z) x^i.
\label{wave-expand}
\ena
Then
\bea
&&
U=\sum_{i=0}^\infty {\mathbb C}\Psi_i^\ast(z).
\label{U-tau}
\ena

By this construction the following property follows.
Let $U$ be a point of UGM,  $\tau({\bf t})$ a tau function corresponding to $U$ and 
$f(z)=e^{\sum_{i=1}^\infty a_i \frac{z^i}{i}}$ an invertible formal power series.
Then $f(z)U$ belongs to UGM and a tau function corresponding to it is given by
\bea
&&
e^{\sum_{i=1}^\infty a_i t_i}\tau({\bf t}).
\label{gauge-trf}
\ena

\section{ (n,1) curve and main results}
In this section the problem is fromulated and main results are stated.




Let $n\geq 2$ be a positive integer and $\alpha$ a complex number.
Consider the rational curve $C_n$ defined by
\bea
&& y^n=x-\alpha_j
\label{Cn}
\ena
which we call $(n,1)$ curve.
The point of infinity of $C_n$ corresponds to $y=\infty$.
Take a local coordinate $z$ around $\infty$ such that
\bea
&&
x=z^{-n},
\hskip10mm
y=z^{-1}(1-\alpha z^n)^{1/n}.
\label{coord-1}
\ena
This type of the local coordinate $z$ was used for a general $(n,s)$ curves of genus $g\geq 1$ 
in constructing the multivariate sigma functions \cite{BEL1997,BEL1999,N2010-1} and the quasi-periodic 
solutions of the $n$-reduced KP-hierarchy\cite{BEL1997, N2010-2}.

Let $V_n$ be the space of meromorphic functions on $C_n$ which have a pole 
only at  $p_\infty$.
It is nothing but the vector space generated by $y^i$, $i\geq 0$:
\bea
&&
V_n=\sum_{i=0}^\infty {\mathbb C} y^i.
\non
\ena
By expanding $y^i$ in the local coordinate $z$ we consider $V_n$ as a subspace of ${\mathbb C}((z))$.
We denote this subspace by $V_n(z)$ indicating the choice of the local coordinate $z$.
Later we consider a coordinate change.
Notice that
\bea
&&
y^i=z^{-i}(1+O(z)), \quad i\geq 0.
\non
\ena
By taking linear cominations of them we have a frame of $V_n(z)$ of the form (\ref{big-cell-1}).
Therefore 
\bea
&&
V_n(z)\in {\rm UGM}^\phi.
\non
\ena

Let $\tau_0({\bf t};V_n(z))$ be the normalized tau function of  $V_n(z)$. 

Our main theorem is

\begin{theorem}\label{main}
The following formula holds:
\bea
&&
\tau_0({\bf t};V_n(z))=e^{\frac{1}{2}\left(q({\bf t})+L({\bf t})\right)},
\non
\\
&&
q({\bf t})=\sum_{i,j\geq 1} q_{i,j}t_it_j,
\hskip5mm
L({\bf t})=\sum_{i=1}^\infty L_{in} t_{in}
\non
\ena
Here, $q_{i,j}=q_{j,i}$ for any $i, j$, $q_{i,j}=0$ if $j=nm$ for some $m\geq 1$ or $i+j\neq0$ ${\rm mod. }n$  and,
 for $r\geq 1$, $s\geq 0$, $1\leq p\leq n-1$, $i\geq 1$,
 \bea
 &&
 q_{nr-p,ns+p}=\alpha^{r+s} \frac{pr}{r+s} \binom{r-\frac{p}{n}}{r}\binom{s+\frac{p}{n}}{s}.
 \label{main-1}
 \ena
 \bea
 &&
 L_{in}=-\sum_{r+s=i, r\geq 1, s\geq 0}\sum_{p=1}^{n-1}
 \frac{ni}{(nr-p)(ns+p)} q_{nr-p,ns+p},
 \label{main-2}
 \ena
 where, for a non-negative integer $n$,
 \bea
 &&
 \binom{x}{n}=\frac{x(x-1)\cdots(x-n+1)}{n!}, \quad n\geq 1,
 \qquad
 \binom{x}{0}=1.
 \non
 \ena
 \end{theorem}
\vskip5mm


The Schur function expansion of $\tau_0({\bf t};V_n(z))$ can also be computed.
\vskip5mm

\begin{theorem}\label{main2}
The following expansion holds.
\bea
&&
\tau_0({\bf t};V_n(z))
\non
\\
&=&
1+
\sum_{l=1}^\infty \sum_{r_i,s_i,p_i,q_i} A_l((r_i),(s_i),(p_i),(q_i))s_{(nr_1-p_1-1,...,nr_l-p_l-1|ns_1+q_1,...,ns_l+q_l)}(t),
\non
\ena
\bea
&&
A_l((r_i),(s_i),(p_i),(q_i))
\non
\\
&&
=
(-1)^{\sum_{i=1}^l(ns_i+q_i+r_i)}\alpha^{\sum_{i=1}^l(r_i+s_i)}
\det\left(\delta_{p_i,q_j}\binom{r_i+s_j-1}{s_j}\binom{s_j+\frac{p_i}{n}}{r_i+s_j}\right)_{1\leq i,j\leq l}.
\non
\ena
Here the sumation in the second term is over all $r_i,s_i,p_i,q_i$ satisfying
\bea
&&
1\leq p_i,q_i\leq n-1,
\quad
r_i\geq 1,
\quad
s_i\geq 0,
\non
\\
&&
nr_1-p_1>\cdots>nr_l-p_l,
\non
\\
&&
ns_1+q_1>\cdots>ns_l+q_l,
\non
\ena
and $(m_1,...,m_l|m_1',...,m_l')$ is the Frobenius notation of a partition.
\end{theorem}

The proofs of  theorems are given in subsequent sections.

\section{Coordinate change and wave functions}
In this section we derive the equation for  $\tau_0({\bf t};V_n(z))$ by considering a coordinate change 
and wave functions.

Let us take $w=y^{-1}$ as another local coordinate around $\infty$.
Then
\bea
&&
x-\alpha=w^{-n},
\hskip10mm
y=w^{-1}.
\label{coord-2}
\ena
By expanding elements of $V_n$ in terms of $w$ and identifying the ambient space $V$ of UGM 
with ${\mathbb C}((w))$, 
we define the subspace $V_n(w)$ of $V$.
 Then
\bea
&&
V_n(w)=\sum_{i=0}^\infty {\mathbb C}w^{-i}=V_\phi\in{\rm UGM}^\phi.
\non
\ena
 So $\tau_0({\bf t};V_n(w))=1$.
Consider the corresponding wave and adjoint wave functions, 
\bea
&&
\Psi(t;w)=e^{\sum_{i=1}^\infty t_i w^{-i}},
\hskip10mm
\Psi^\ast(t;w)=e^{-\sum_{i=1}^\infty t_i w^{-i}}.
\label{wave-w}
\ena

Here we change the local coordinate from $w$ to $z$.
By (\ref{coord-2}) $z$ and $w$ are connected by
\bea
&&
z^{-n}-\alpha=w^{-n}.
\non
\ena
Therefore
\bea
&&
w=z(1-\alpha z^n)^{-1/n}.
\label{c-change}
\ena
Expand 
\bea
&&
(1-\alpha z^n)^{i/n}=\sum_{m=0}^\infty a_{i,m} z^m,
\label{expansion}
\ena
where
\bea
&&
a_{i,m}=
\left\{
\begin{array}{rl}
(-\alpha)^k\binom{\frac{i}{n}}{k}, &m=nk \text{ for some $k\geq 0$}\\
0, & \text{otherwise}.
\end{array}
\right.
\label{aim}
\ena
We set $a_{i,j}=0$ if $j<0$ for the sake of convenience.
Let us decompose $w^{-i}$ into two parts corresponding to negative and non-negative powers in $z$,
\bea
&&
w^{-i}=\sum_{m=0}^{i-1}a_{i,m}z^{-(i-m)}+f_i(z),
\quad
f_i(z)=\sum_{m=i}^\infty a_{i,m} z^{m-i}.
\label{fi}
\ena
Then 
\bea
\sum_{i=1}^\infty t_i w^{-i}=
\sum_{i=1}^\infty t_i\sum_{m=0}^{i-1}a_{i,m}z^{-(i-m)}+\sum_{i=1}^\infty t_i f_i(z).
\non
\ena
Define the new set of time variables $\{T_i| i\geq 1\}$ by
\bea
&&
T_i=\sum_{j=i}^\infty a_{j,j-i}  t_j.
\label{Ti}
\ena
Then 
\bea
&&
\sum_{i=1}^\infty t_i w^{i}=\sum_{l=1}^\infty T_l z^{-l}+\sum_{i=1}^\infty t_i f_i(z).
\label{tTf}
\ena
The coefficient matrix $(a_{j,j-i})_{i,j\geq 1}$ in the right hand side of  (\ref{Ti})  is an upper triangular matrix whose diagonal entries 
are all one. So it has the inverse which is again an upper triangular matrix with the same property.
Therefore $t_i$ can be written as
\bea
&&
t_i=\sum_{j=1}^\infty b_{i,j}T_j,
\label{ti}
\ena
where $b_{i,j}=0$ if $j<i$ and $b_{i,i}=1$ for $i\geq 1$.
Then 
\bea
&&
\sum_{i=1}^\infty t_i f_i(z)
=\sum_{j=1}^\infty T_j F_j(z),
\label{tTF}
\ena
with
\bea
&&
F_j(z)=\sum_{m=0}^\infty \left(\sum_{i=1}^j b_{i,j}a_{i,m+i}\right) z^m.
\label{Fj}
\ena
Substituting (\ref{tTf}) and (\ref{tTF}) into (\ref{wave-w}) we get
\bea
&&
\Psi({\bf t};w)=e^{\sum_{i=1}^\infty T_i F_i(z)}e^{\sum_{i=1}^\infty T_j z^{-j}},
\hskip5mm
\Psi^\ast({\bf t};w)=e^{-\sum_{i=1}^\infty T_i F_i(z)}e^{-\sum_{i=1}^\infty T_j z^{-j}}.
\non
\ena
Consider $w$ as a function of $z$ by (\ref{c-change}), $w=w(z)$.
Define new pair of functions of ${\bf T}=(T_i)$ and $z$ by
\bea
\tilde{\Psi}({\bf T};z)&=&
\frac{z^2}{w^2}\frac{d w}{d z}e^{\sum_{i=1}^\infty T_i F_i^{+}(z)}e^{\sum_{i=1}^\infty T_j z^{-j}},
\label{wave-z1}
\\
\tilde{\Psi}^\ast({\bf T};z)&=&e^{-\sum_{i=1}^\infty T_i F_i^{+}(z)}e^{-\sum_{i=1}^\infty T_j z^{-j}},
\label{wave-z2}
\ena
where 
\bea
&&
F_i^+(z)=F_i(z)-F_i(0).
\non
\ena

Then

\begin{lemma} The functions $\tilde{\Psi}({\bf T};z)$ and $\tilde{\Psi}^\ast({\bf T};z)$ satisfy, for any 
${\bf T}=(T_i)$ and 
${\bf T}'=(T_i')$, the following bilinear equation,
\bea
&&
{\rm Res}_{z=0}\tilde{\Psi}({\bf T};z)\tilde{\Psi}^\ast({\bf T}';z)\frac{dz}{z^2}=0.
\label{bilinear1}
\ena
\end{lemma}
\vskip1mm
\noindent
{\it Proof.} Obviously we have
\bea
&&
{\rm Res}_{w=0}\Psi({\bf t};w)\tilde\Psi^\ast({\bf t};w)\frac{dw}{w^2}=0.
\non
\ena
Changing the variable from $w$ to $z$ and multiplying by $e^{\sum_{i=1}^\infty (-T_i+T_i')F_i(0)}$
we get (\ref{bilinear1}). \qed

Recall the following characterization of the KP-hierarchy in terms of wave functions.

\begin{theorem}\label{djkm}\cite{JM,DJKM}
Suppose we have formal series of the form
\bea
\Phi({\bf t};z)&=&(1+\sum_{j=1}^\infty \Phi_j({\bf t}) z^j)e^{\sum_{i=1}^\infty t_i z^{-i}},
\label{singular-1}
\\
\Phi^\ast({\bf t};z)&=&(1+\sum_{j=1}^\infty \Phi_j^\ast({\bf t}) z^j)e^{-\sum_{i=1}^\infty t_i z^{-i}},
\label{singular-2}
\ena
which satisfy 
\bea
&&
{\rm Res}_{z=0} \Phi({\bf t};z)\Phi^\ast({\bf t}';z)\frac{dz}{z^2}=0,
\non
\ena
for any $t$ and $t'$.
Then there exists a solution $\tau({\bf t})$ of the KP-hierarchy, unique up to constant multiples,
 such that $\Phi({\bf t};z)$ and $\Phi^\ast({\bf t};z)$ are the corresponding wave and adjoint wave functions.
\end{theorem}

By   (\ref{c-change}) , (\ref{wave-z1}), (\ref{wave-z2})  it is obvious that 
$\tilde{\Psi}({\bf T};z)$ and $\tilde{\Psi}^\ast({\bf T};z)$ have the expansions 
of the form (\ref{singular-1}) and (\ref{singular-2}) respectively.
Therefore 

\begin{corollary}\label{existence-tau}
There exists a unique, up to constant multiples, a solution $\tau({\bf T})$ of the KP-hierarchy with the 
time variables ${\bf T}=(T_i)$ such that
\bea
\tilde{\Psi}({\bf T};z)&=&\frac{\tau({\bf T}-[z])}{\tau({\bf T})} e^{\sum_{i=1}^\infty T_i z^{-i}},
\label{existence-tau1}
\\
\tilde{\Psi}^\ast({\bf T};z)&=&\frac{\tau({\bf T}+[z])}{\tau({\bf T})}e^{-\sum_{i=1}^\infty T_i z^{-i}}.
\label{existence-tau2}
\ena
\end{corollary}

Next we prove

\begin{prop}\label{point-UGM}
The point of UGM corresponding to $\tau({\bf T})$ in Corollary \ref{existence-tau} is $V_n(z)$.
\end{prop}
\vskip1mm
\noindent
{\it Proof.} We prove the proposition by calculating
 the expansion of $\tilde{\Psi}^\ast(T_1,0,0,...;z)$ in $T_1$.
Let 
\bea
&&
g(z)=w^{-1}=z^{-1}+\sum_{m=0}^\infty a_{1,m+1}z^m.
\non
\ena
Then, by the definition,
\bea
&&
V_n(z)=\sum_{i=0}^\infty {\mathbb C} g(z)^i.
\non
\ena
On the other hand 
\bea
&&
\tilde{\Psi}^\ast(T_1,0,0,...;z)=e^{-T_1(z^{-1}+F_1^+(z))}.
\non
\ena
We have 
\bea
&&
F_1(z)=\sum_{m=0}^\infty a_{1,m+1} z^m,
\non
\ena
since $b_{1,1}=1$. 
Therefore 
\bea
&&
\tilde{\Psi}^\ast(T_1,0,0,...;z)=e^{-T_1(g(z)-a_{1,1})}.
\non
\ena
It follows that 
\bea
&&
{\rm Span}_{{\mathbb C}}\left\{\partial_{T_1}^i\tilde{\Psi}(T_1,0,0,...;z)|_{T_1=0} | i\geq 0\right\}
=\sum_{i=0}^\infty {\mathbb C} g(z)^i =V_n(z),
\non
\ena
which completes the proof of the proposition. \qed

\begin{corollary}\label{upto-constant} The tau function $\tau({\bf T})$ in Corollary \ref{existence-tau}
is a non-zero constant multiple of $\tau_0({\bf T};V_n(z))$.
\end{corollary}

\section{Expansion coefficients of $F_i^+(z)$}
In this section we compute the expansion coefficients of $F_i^+(z)$ which appears in 
the wave and the adjoint wave functions  (\ref{wave-z1}), (\ref{wave-z2}).

Set
\bea
&&
-F_i^{+}(z)=\sum_{j=1}^\infty c_{i,j}\frac{z^j}{j}.
\label{cij-1}
\ena

\begin{prop}\label{cij-2} The following properties hold.
\vskip2mm
\noindent
(i) $c_{i,j}=c_{j,i}$ 
\vskip2mm
\noindent
(ii) $c_{i,j}=0$ if $i+j\neq 0$ mod. $n$.
\vskip2mm
\noindent
(iii) For $(i,j)=(ns+p,nr-p)$ with $r\geq 1$, $s\geq 0$, $1\leq p\leq n$, 
\bea
&&
 c_{i,j}=\alpha^{r+s} \frac{pr}{r+s} \binom{r-\frac{p}{n}}{r}\binom{s+\frac{p}{n}}{s}.
 \label{cij-form}
 \ena
 \vskip2mm
 \noindent
 (iv) If $j=0$ mod.$n$, $c_{i,j}=0$.
 \end{prop}

By the definition (\ref{Fj}) of $F_i(z)$ 
\bea
&&
c_{i,j}=-j\sum_{l=1}^i b_{l,i} a_{l,j+l}.
\label{cij-3}
\ena
So we need to compute $(b_{i,j})=(a_{j,j-i})^{-1}$.

\begin{lemma}\label{bij-expr} The following properties are valid.
\vskip2mm
\noindent
(i) $b_{i,j}=0$ if $i\neq j$ mod. $n$.
\vskip2mm
\noindent
(ii) For $r,s\geq 0$, $1\leq p\leq n$,
\bea
&&
b_{nr+p,ns+p}=\alpha^{s-r}\binom{s+\frac{p}{n}}{s-r}.
\label{bij}
\ena
\end{lemma}
\vskip1mm
\noindent
{\it Proof.}
Since $a_{i,j}=0$ for $j\neq 0$ mod. $n$, the equation (\ref{Ti}) can be written as 
\bea
&&
T_{nr+p}=\sum_{s=r}^\infty a_{ns+p,n(s-r)}t_{ns+p},
\label{pr1}
\ena
with $r,s\geq 0$, $1\leq p\leq n$.
It follows that $b_{i,j}=0$ unless $i=j$ mod. $n$. Then (\ref{ti}) reduces to
\bea
&&
t_{nr+p}=\sum_{s=0}^\infty b_{nr+p,ns+p}T_{ns+p}.
\label{pr2}
\ena
So it is sufficient to prove that $\{b_{i,j}\}$ given by (\ref{bij}) satisfy, for any $r,r'\geq 0$, 
\bea
&&
\sum_{s=r}^{r'} a_{ns+p,n(s-r)}b_{ns+p,nr'+p}=\delta_{r,r'}.
\label{pr3}
\ena
By the definition (\ref{aim}) of $a_{i,j}$
\bea
&&
a_{ns+p,n(s-r)}=(-\alpha)^{s-r}\binom{s+\frac{p}{n}}{s-r}.
\label{pr4}
\ena
Substituting (\ref{bij}) and (\ref{pr4}) into (\ref{pr3}) we have
\bea
&&
\alpha^{r'-r}\sum_{s=r}^{r'} (-1)^{s-r}\binom{s+\frac{p}{n}}{s-r}\binom{r'+\frac{p}{n}}{r'-s}
=\delta_{r,r'}.
\label{pr5}
\ena
Let us prove this equation.
By computation we have
\bea
&&
\binom{s+\frac{p}{n}}{s-r}\binom{r'+\frac{p}{n}}{r'-s}
=
\binom{r'+\frac{p}{n}}{r'-r}\binom{r'-r}{r'-s}.
\non
\ena
Then 
\bea
\text{LHS of (\ref{pr5})}&=&
\alpha^{r'-r}\binom{r'+\frac{p}{n}}{r'-r}\sum_{s=r}^{r'}(-1)^{s-r}\binom{r'-r}{r'-s}
\non
\\
&=&
\alpha^{r'-r}\binom{r'+\frac{p}{n}}{r'-r}\sum_{s=0}^{r'-r}(-1)^{s}\binom{r'-r}{s}
\non
\\
&=&\delta_{r,r'}.
\non
\ena
 \qed

\noindent
{\it Proof of Proposition \ref{cij-2}}\par

Notice that, by (\ref{cij-3}), if $c_{i,j}\neq 0$ then there exists $l$ such that 
$l=i$, $j+l=0$ mod. $n$. Then $i=l=-j$ mod. $n$. So $c_{i,j}=0$ if $-i\neq j$ mod. $n$.
This proves (ii).

For $i=ns+p$, $j=nr-p$, $s\geq 0$, $r\geq 1$, $1\leq p\leq n$, we have, using Lemma \ref{bij-expr},
\bea
c_{ns+p,nr-p}&=&-(nr-p)\sum_{l=1}^{ns+p} b_{l,ns+p} a_{l, nr-p+l}
\non
\\
&=&
-(nr-p)\sum_{k=0}^{s} b_{nk+p,ns+p} a_{nk+p, n(r+k)}
\non
\\
&=&
-(nr-p)\alpha^{r+s}\sum_{k=0}^s (-1)^{r+k}\binom{s+\frac{p}{n}}{s-k}\binom{k+\frac{p}{n}}{r+k}.
\label{cij-prf1}
\ena
By computation we have
\bea
&&
(nr-p)\binom{s+\frac{p}{n}}{s-k}\binom{k+\frac{p}{n}}{r+k}=
(-1)^{r-1}p\binom{s+\frac{p}{n}}{s}\binom{r-\frac{p}{n}}{r}
\binom{r+s}{s}^{-1}\binom{r+s}{s-k}.
\non
\ena
Then 
\bea
c_{ns+p,nr-p}&=&
 \alpha^{r+s}p
\binom{s+\frac{p}{n}}{s}\binom{r-\frac{p}{n}}{r}
\binom{r+s}{s}^{-1}
\sum_{k=0}^s (-1)^{k} \binom{r+s}{s-k},
\non
\\
&=&
(-1)^s\alpha^{r+s}p
\binom{s+\frac{p}{n}}{s}\binom{r-\frac{p}{n}}{r}
\binom{r+s}{s}^{-1}
\sum_{k=0}^s (-1)^{k} \binom{r+s}{k}.
\label{cij-prf2}
\ena

\begin{lemma}\label{sum-1} The following formula holds,
\bea
&&
\sum_{k=0}^s (-1)^{k}\binom{r+s}{k}=(-1)^s\binom{r+s-1}{s}.
\non
\ena
\end{lemma}
\vskip1mm
\noindent
{\it Proof.}
We have
\bea
\sum_{k=0}^s (-1)^{k}\binom{r+s}{k}&=&
\sum_{k=0}^{s}(-1)^k\left(
\binom{r+s-1}{k}+\binom{r+s-1}{k-1}
\right)
\non
\\
&=&
\sum_{k=0}^{s-1}(-1)^k\binom{r+s-1}{k}+(-1)^s\binom{r+s-1}{s}+\sum_{k=1}^{s}(-1)^k\binom{r+s-1}{k-1}
\non
\\
&=& 
(-1)^s\binom{r+s-1}{s}.
\non
\ena
 \qed

Applying Lemma \ref{sum-1} to (\ref{cij-prf2}) we get (iii) of the proposition.

If $p=n$ the right hand side of (\ref{cij-form}) is zero , since
\bea
&&
\binom{r-\frac{p}{n}}{r}=0.
\non
\ena
So $c_{i,j}=0$ if $j=0$ mod. $n$. Thus (iv) of the proposition is proved.

Let us prove (i) of the proposition.
It is sufficient to prove $c_{ns+p,nr-p}=c_{nr-p,ns+p}$, $r\geq 1$, $s\geq 0$, $1\leq p\leq n-1$.
Using (\ref{cij-form}) we have
\bea
c_{nr-p,ns+p}
&=&c_{n(r-1)+n-p, n(s+1)-(n-p)}
\non
\\
&=&
\alpha^{r+s}\frac{(s+1)(n-p)}{r+s}\binom{r-1+\frac{n-p}{n}}{r-1}\binom{s+1-\frac{n-p}{n}}{s+1}
\non
\\
&=&
\alpha^{r+s}\frac{(s+1)(n-p)}{r+s}\binom{r-\frac{p}{n}}{r-1}\binom{s+\frac{p}{n}}{s+1}.
\non
\ena
Using 
\bea
(n-p)\binom{r-\frac{p}{n}}{r-1}&=&rn\binom{r-\frac{p}{n}}{r},
\non
\\
\binom{s+\frac{p}{n}}{s+1}&=&\frac{p}{n(s+1)}\binom{s+\frac{p}{n}}{s},
\non
\ena
we get
\bea
c_{nr-p,ns+p}&=&
\alpha^{r+s}\frac{pr}{r+s}\binom{r-\frac{p}{n}}{r}\binom{s+\frac{p}{n}}{s}.
\non
\\
&=&
c_{ns+p,nr-p}.
\non
\ena
\qed

\section{Proof of Theorem \ref{main}}

We shall find the function $\tau({\bf T})$ which satisfies (\ref{existence-tau1})and 
(\ref{existence-tau2}) in the form
\bea
&&
\tau({\bf T})=e^{\frac{1}{2}(q({\bf T}|{\bf T})+L({\bf T}))},
\label{ansatz1}
\\
&&
q({\bf T}|{\bf S})=\sum_{i,j=1}^\infty q_{i,j}T_iS_j,
\hskip5mm
L({\bf T})=\sum_{i=1}^\infty L_i T_i,
\label{ansatz2}
\ena
where $q_{i,j}=q_{j,i}$ for any $i,j$, ${\bf S}=(S_1,S_2,...)$.
 The equation (\ref{existence-tau2}) is equivalent to
\bea
&&
\frac{\tau({\bf T}+[z])}{\tau({\bf T})}=
e^{q({\bf T}|[z])+\frac{1}{2}(q([z]|[z])+L([z]))}
\non
\ena
which, by (\ref{ansatz1}), (\ref{ansatz2}),  is rewritten as
\bea
&&
q({\bf T}|[z])+\frac{1}{2}\left(q([z]|[z])+L([z])\right)=\sum_{i,j=1}^\infty c_{i,j}T_i\frac{z^j}{j}.
\label{tau-1}
\ena
This equation is satisfied if we set
\bea
q({\bf T}|[z])&=&\sum_{i,j=1}^\infty c_{i,j}T_i\frac{z^j}{j},
\label{tau-2}
\\
q([z]|[z])+L([z])&=&0.
\label{tau-3}
\ena
These equations are solved if we take 
\bea
q_{i,j}&=&c_{i,j},
\label{tau-4}
\\
L_k&=&-\sum_{i+j=k}\frac{k}{ij}c_{i,j}.
\label{tau-5}
\ena

Now Theorem \ref{main} follows from the following lemma which is used in the proof of 
Theorem \ref{djkm} in \cite{DJKM}.

\begin{lemma}{\rm \cite{DJKM}}\label{tau-unique}
Let $\tau_i({\bf t})$, $i=1,2$, be two functions of ${\bf t}=(t_1,t_2,...)$ such that
\bea
&&
\frac{\tau_1({\bf t}+[z])}{\tau_1({\bf t})}=\frac{\tau_2({\bf t}+[z])}{\tau_2({\bf t})}
\non
\ena
Then $\tau_2({\bf t})=c\tau_1({\bf t})$ for some constant $c$.
\end{lemma}
\vskip1mm
\noindent
{\it Proof.} Let $f({\bf t})=\tau_2({\bf t})/\tau_1({\bf t})$. Then the condition 
is 
\bea
&&
f({\bf t}+[z])=f({\bf t}).
\label{shift1}
\ena
Using 
\bea
&&
f({\bf t}+[z])=e^{\sum_{i=1}^\infty \frac{z^i}{i}\partial_i} f({\bf t})
=\sum_{i=0}^\infty z^i p_i(\tilde{\partial}) f({\bf t}),
\label{shift2}
\ena
where $ \partial_i=\partial/\partial t_i$ and $\tilde{\partial}=(\partial_1,\partial_2/2,\partial_3/3,...)$,
we have 
\bea
&&
\sum_{i=1}^\infty z^i  p_i(\tilde{\partial})f({\bf t})=0.
\label{shift3}
\ena
Therefore
\bea
&&
p_i(\tilde{\partial})f({\bf t})=0, \quad i\geq 1.
\label{shift4}
\ena
Since 
\bea
&&
p_i(\tilde{\partial})=\frac{\partial_i}{i}+(\text{terms containing only $\partial_j$, $j<i$}),
\non
\ena
we get $\partial_i f(t)=0$ for any $i\geq 1$. \qed

By Corollary \ref{upto-constant} both $\tau({\bf T})$ constructed above and $\tau_0({\bf T};V_n(z))$ saisfy 
the equation (\ref{existence-tau2}). 
Therefore $\tau({\bf T})=c\tau_0({\bf T};V_n(z))$ for some constant $c$ by Lemma \ref{tau-unique}.  
By setting $T_j=0$ for all $j$ we have $c=1$. 
Then we have (\ref{main-1}),  (\ref{main-2}) by (\ref{tau-4}), (\ref{tau-5}), Proposition \ref{cij-2}.
Thus Theorem \ref{main} is proved. \qed

\section{Proof of Theorem \ref{main2}}

By the definition of $V_n(z)$ the following set of functions give a basis of $V_n(z)$,
\bea
h_{ni}(z):&=&z^{-ni},\quad i\geq 0,
\non
\\
h_i(z):&=&y^i=z^{-i}(1-\alpha z^n)^{\frac{i}{n}}, \quad i\geq 1,\quad  i\neq 0 \text{ mod.}n.
\non
\ena
Then it is obvious that there exists the unique basis $\tilde{h}_i$, $i\geq 1$ with the property
\bea
&&
\tilde{h}_i=z^{-i}+O(z)
\non
\ena
where $O(z)$ denotes an element in $z{\mathbb C}[[z]]$.

It is given explicitly as

\begin{prop}\label{S1}
The basis $\{\tilde{h}_i(z)\}$ is given by
\bea
\tilde{h}_{nr}(z)&=&z^{-nr},
\non
\\
\tilde{h}_{nr+p}(z)
&=&
z^{-nr-p}+\sum_{k=1}^\infty (-1)^k \alpha^{k+r}\binom{k+r-1}{r}\binom{r+\frac{p}{n}}{k+r}
z^{nk-p},
\label{S-1}
\ena
where $r\geq 0$, $1\leq p\leq n-1$.
\end{prop}
\vskip1mm
\noindent
{\it Proof.}
It is sufficient to prove that $\tilde{h}_{nr+p}(z)$ is a linear combination 
of $h_{nr'+p}(z)$, $r'\geq 0$ for each $p$.
We prove it by induction on $r$.

By (\ref{expansion}), (\ref{aim}) 
\bea
&&
h_{nr+p}(z)=z^{-nr-p}+\sum_{k=1}^\infty a_{nr+p,nk} z^{n(k-r)-p}.
\non
\ena
For $r=0$ we have
\bea
&&
h_p(z)=z^{-p}+\sum_{k=1}^\infty (-\alpha)^k\binom{\frac{p}{n}}{k}z^{nk-p}=\tilde{h}_p(z).
\non
\ena
Suppose that the assertion is valid until $r$. 
Dividing $h_{n(r+1)+p}(z)$ to negative and non-negative power parts in $z$ we have
\bea
h_{n(r+1)+p}(z)&=&z^{-n(r+1)-p}+\sum_{l=1}^{r+1}a_{n(r+1)+p,nl} z^{-n(r+1-l)-p}
\non
\\
&&
+\sum_{k=1}^\infty a_{n(r+1)+p,n(k+r+1)}z^{nk-p}.
\label{S-2}
\ena
We erase the middle trem by subtracting linear combination of 
$\tilde{h}_{nk+p}(z)$, $0\leq k\leq r$. Set
\bea
&&
h'_{n(r+1)+p}(z)=h_{n(r+1)+p}(z)-\sum_{l=1}^{r+1}a_{n(r+1)+p,nl} \tilde{h}_{n(r+1-l)+p}(z).
\non
\ena
We shall show $h'_{n(r+1)+p}(z)=\tilde{h}_{n(r+1)+p}(z)$. 

Using the assumption of induction we have
\bea
&&
h'_{n(r+1)+p}(z)
\non
\\
&=&
z^{-n(r+1)-p}
+
\sum_{k=1}^\infty
\biggl(
a_{n(r+1)+p,n(k+r+1)}
-\sum_{l=1}^{r+1}(-1)^k a_{n(r+1)+p,nl}\alpha^{k+r+1-l}
\non
\\
&&
\hskip40mm
\times\binom{k+r-l}{r+1-l}\binom{r+1-l+\frac{p}{n}}{k+r+1-l}
\biggr)
z^{nk-p}.
\label{S-3}
\ena
By (\ref{aim}) the coefficient of $z^{nk-p}$ of the right hand side 
is
\bea
&&
(-\alpha)^{k+r+1}
\biggl(
\binom{r+1+\frac{p}{n}}{r+1+k}
\non
\\
&&
-\sum_{l=1}^{r+1}(-1)^{l+r+1}
\binom{r+1+\frac{p}{n}}{l}
\binom{r+1-l+\frac{p}{n}}{k+r+1-l}
\binom{k+r-l}{r+1-l}
\biggr).
\label{S-4}
\ena
Notice the equation
\bea
&&
\binom{r+1+\frac{p}{n}}{l}
\binom{r+1-l+\frac{p}{n}}{k+r+1-l}
=
\binom{r+1+\frac{p}{n}}{k+r+1}
\binom{k+r+1}{l}.
\non
\ena
Then 
\bea
&&
\text{the right hand side of (\ref{S-4})}
\non
\\
&=&
(-\alpha)^{k+r+1}\binom{r+1+\frac{p}{n}}{r+1+k}
\biggl(
1-\sum_{l=0}^r
(-1)^l\binom{k+r+1}{r+1-l}
\binom{k-1+l}{l}
\biggr).
\non
\ena

\begin{lemma}\label{S2}
The following equation holds:
\bea
&&
1-\sum_{l=0}^r
(-1)^l\binom{k+r+1}{r+1-l}
\binom{k-1+l}{l}
=(-1)^{r+1}\binom{k+r}{r+1}.
\non
\ena
\end{lemma}
\vskip1mm
\noindent
{\it Proof.}
It is sufficient to prove the following identity for polynomials in $x$:
\bea
&&
1-\sum_{l=0}^r
(-1)^l\binom{x+r+1}{r+1-l}
\binom{x-1+l}{l}
=(-1)^{r+1}\binom{x+r}{r+1}.
\non
\ena
It is easily proved by examining the zeros and the coefficients of $x^{r+1}$ of both sides.
We leave the details to the reader. \qed
\vskip5mm

By Lemma \ref{S2} we finally have 
\bea
&&
h'_{n(r+1)+p}(z)
=
z^{-n(r+1)-p}+
\sum_{k=1}^\infty
(-1)^k\alpha^{k+r+1}
\binom{r+1+\frac{p}{n}}{r+1+k}
\binom{k+r}{r+1}z^{nk-p}
\non
\ena
which is equal to $\tilde{h}_{n(r+1)+p}(z)$. This completes the proof 
of Proposition \ref{S1}.
\qed

By Proposition \ref{S1}
\bea
&&
\xi=[...,\tilde{h}_1(z),\tilde{h}_0(z)]
\non
\ena
is a frame of $V_n(z)$ of the form (\ref{big-cell}).
To determine the Schur function expansion of $\tau_0({\bf t};V_n(z))$ we have to compute 
the Pl\"ucker coordinates of $\xi$.
The Pl\"ucker coordinates corresponding to hook diagrams is easily computed as

\begin{lemma}\label{S3} 
Let $\xi'=[...,\xi'_{-2}, \xi'_{-1},\xi'_0]$ be a frame of a point of UGM of the form (\ref{big-cell}), that is, 
\bea
&&
\xi'_j=z^{j}+\sum_{i=1}^\infty x_{i,j} z^i,
\quad
j\leq 0.
\non
\ena
Then, for $i, j\geq 0$, 
\bea
&&
\xi'_{(i|j)}=(-1)^j x_{i+1,-j}.
\non
\ena
\end{lemma}

By this lemma and Proposition \ref{S1} we get

\begin{corollary}\label{S4}
(i) $\xi_{(i|j)}=0$ if $j=0$ mod.$n$ or $i+j+1\neq 0$ mod.$n$.
\vskip2mm
\noindent
(ii) For $r\geq 1$, $s\geq 0$, $1\leq p, q\leq n-1$, 
\bea
&&
\xi_{(nr-p-1|ns+q)}=
\delta_{p,q}(-1)^{ns+q+r}
\binom{r+s-1}{s}
\binom{s+\frac{p}{n}}{r+s} \alpha^{r+s}.
\non
\ena
\end{corollary}

Finally let us recall the Giambelli formula for Pl\"ucker coordinates.

\begin{theorem}\label{S5}\cite{EH,NOS}
Let $\xi'$ be the frame of a point of ${\rm UGM}$ which is the same form as in Lemma \ref{S3}. Then 
for $l\geq 1$, $a_1>\cdots>a_l\geq 0$, $b_1>\cdots>b_l\geq 0$, 
\bea
&&
\xi'_{(a_1,...,a_l|b_1,...,b_l)}=\det\left(\xi'_{(a_i|b_j)}\right)_{1\leq i,j\leq l}.
\non
\ena
\end{theorem}

Now Theorem \ref{main2} follows from Theorem \ref{S5} and Corollary \ref{S4}.
\qed

\section{Generating function of $q_{i,j}$}
In this section we calculate the generating function of $\{q_{i,j}\}$ which is used to compute the action 
of vertex operators on $\tau_0({\bf t};V_n(z))$.

Let
\bea
&&
G(z)=(1-\alpha z^n)^{\frac{1}{n}},
\qquad
H(z_1,z_2)=z_2G(z_1)-z_1G(z_2).
\label{H}
\ena
Since $H(z_1,z_1)=0$, $H(z_1,z_2)/(z_2-z_1)$ is holomorphic near $(0,0)$. 
It has the expansion of the form
\bea
&&
\frac{H(z_1,z_2)}{z_2-z_1}=1+\text{h.o.t},
\non
\ena
where h.o.t means terms containing $z_1^iz_2^j$ with $i+j\geq 1$. 
Thus $\log(H(z_1,z_2)/(z_2-z_1))$ can be expanded to a power series in $z_1,z_2$.

\begin{prop}\label{bidif}
The following expansion holds,
\bea
&&
\partial_{z_1}\partial_{z_2}\log\left(\frac{H(z_1,z_2)}{z_2-z_1}\right)=
\sum_{i,j=1}^\infty q_{i,j}z_1^{i-1}z_2^{j-1},
\quad
\partial_{z_i}=\partial/\partial z_i.
\label{bidif-1}
\ena
\end{prop}
\vskip2mm

\begin{remark} For $(p_1,p_2)\in C_n\times C_n$, $p_i=(x_i,y_i)$, the bilinear differential 
\bea
&&
d_{z_1}d_{z_2}\log H(z_1,z_2)=d_{p_1}d_{p_2}\log(y_1-y_2)=
d_{p_2}\frac{\sum_{j=0}^{n-1}y_1^jy_2^{n-j}}{(x_1-x_2)ny_1^{n-1}}dx_1
\non
\ena
is the genus zero analogue of  that of an $(n,s)$ curve with positive genus 
which is used in the construction of the multi-variate sigma function 
\cite{BEL1997,BEL1999, N2010-1}.
\end{remark}
\vskip3mm
\noindent
{\it Proof of Proposition \ref{bidif}.}\par
\vskip2mm
We have
\bea
&&
\partial_{z_1}\partial_{z_2}\log\left( \frac{H(z_1,z_2)}{z_2-z_1} \right)
=\partial_{z_1}\partial_{z_2}
\log\left( 1-\frac{z_1}{z_2}\frac{G(z_2)}{G(z_1)} \right)-\frac{1}{(z_2-z_1)^2}.
\label{bidif-2}
\ena
Let us expand the right hand side near $(z_1,z_2)=(0,0)$ in $\mathbb{C}^2$ assuming 
\bea
&&
\left|\frac{z_1}{z_2}\right|<1,
\quad
\left|\frac{z_1}{z_2}\frac{G(z_2)}{G(z_1)}\right|<1,
\quad
z_1z_2\neq 0.
\non
\ena
For example 
\bea
&&
\frac{1}{(z_2-z_1)^2}=\sum_{m=1}^\infty m z_1^{m-1}z_2^{-m-1}.
\non
\ena
Since the left hand side of (\ref{bidif-2}) is holomorphic at $(0,0)$, only non-negative powers 
of $z_1, z_2$ should remain in the expansion of the right hand side under the assumption above.
Define the symbol $[\quad]_+$ by
\bea
&&
\left[ \sum_{i,j=-\infty}^{\infty} e_{i,j} z_1^i z_2^j\right]_+
=\sum_{i,j=0}^\infty e_{i,j} z_1^i z_2^j.
\non
\ena
Then 
\bea
&&
\partial_{z_1}\partial_{z_2}\log\left( \frac{H(z_1,z_2)}{z_2-z_1} \right)
=\left[\partial_{z_1}\partial_{z_2}
\log\left( 1-\frac{z_1}{z_2}\frac{G(z_2)}{G(z_1)} \right)\right]_+.
\label{bidif-3}
\ena
Write
\bea
&&
\left[\partial_{z_1}\partial_{z_2}\log\left( 1-\frac{z_1}{z_2}\frac{G(z_2)}{G(z_1)} \right)\right]_+
=\sum_{i,j=1}^\infty \tilde{q}_{i,j} z_1^{i-1}z_2^{j-1}.
\ena
We shall show that $\tilde{q}_{i,j}=q_{i,j}$.

Expand the logarithmic function as
\bea
&&
\log\left( 1-\frac{z_1}{z_2}\frac{G(z_2)}{G(z_1)} \right)
=-\sum_{m=1}^\infty \frac{1}{m}\left(\frac{G(z_2)}{G(z_1)}\right)^m\left(\frac{z_1}{z_2}\right)^m.
\label{bidif-4}
\ena
We further expand 
\bea
&&
\left(\frac{G(z_2)}{G(z_1)}\right)^m
=(1-\alpha z_1^n)^{-\frac{m}{n}}(1-\alpha z_2^n)^{\frac{m}{n}}
=
\sum_{k,l=0}^\infty (-\alpha)^{k+l}\binom{-\frac{m}{n}}{k}\binom{\frac{m}{n}}{l}z_1^{nk}z_2^{nl}.
\label{bidif-5}
\ena
Substituting this expression into (\ref{bidif-4}) and writing $m=ni+p$, $i\geq 0$, $1\leq p\leq n$,
we get
\bea
&&
\text{RHS of (\ref{bidif-4})}
\non
\\
&=&
-\sum_{p=1}^n\sum_{i=0}^\infty \frac{1}{ni+p}
\sum_{k,l=0}^\infty 
(-\alpha)^{k+l}\binom{-i-\frac{p}{n}}{k}\binom{i+\frac{p}{n}}{l}z_1^{n(k+i)+p}z_2^{n(l-i)-p}
\non
\\
&=&
-\sum_{p=1}^n\sum_{i=0}^\infty \frac{1}{ni+p}
\sum_{s\geq i, r\geq -i}^\infty 
(-\alpha)^{r+s}\binom{-i-\frac{p}{n}}{s-i}\binom{i+\frac{p}{n}}{r+i}z_1^{ns+p}z_2^{nr-p}.
\non
\ena
Then
\bea
&&
\left[
\partial_{z_1}\partial_{z_2}\log\left( 1-\frac{z_1}{z_2}\frac{G(z_2)}{G(z_1)} \right) 
\right]_+
\non
\\
&=&
\sum_{p=1}^n\sum_{r\geq 1,s\geq 0}\tilde{q}_{ns+p,nr-p} z_1^{ns+p-1}z_2^{nr-p-1},
\non
\ena
where
\bea
&&
\tilde{q}_{ns+p,nr-p}
=
(-1)^{r+s+1}\alpha^{r+s}
\sum_{i=0}^s \frac{(ns+p)(nr-p)}{ni+p} \binom{-i-\frac{p}{n}}{s-i}\binom{i+\frac{p}{n}}{r+i}.
\non
\ena
By this expression we see that $\tilde{q}_{i,j}=0$ if $i+j\neq 0$ mod. $n$.

A computation shows 
\bea
&&
(ns+p)(nr-p)\binom{-i-\frac{p}{n}}{s-i}\binom{i+\frac{p}{n}}{r+i}
\non
\\
&=&
(-1)^{r+s+i-1}p(ni+p)\frac{r!s!}{(r+i)!(s-i)!}\binom{s+\frac{p}{n}}{s}\binom{r-\frac{p}{n}}{r}.
\non
\ena
Then 
\bea
&&
\tilde{q}_{ns+p,nr-p}=
pr!s! \alpha^{r+s}
\binom{s+\frac{p}{n}}{s}\binom{r-\frac{p}{n}}{r}
\sum_{i=0}^s \frac{(-1)^i}{(r+i)!(s-i)!}.
\label{bidif-6}
\ena
Using Lemma \ref{sum-1} we easily have
\bea
&&
\sum_{i=0}^s \frac{(-1)^i}{(r+i)!(s-i)!}=\frac{1}{(r+s)(r-1)!s!}.
\non
\ena
Substituting it to (\ref{bidif-6}) we get 
\bea
&&
\tilde{q}_{ns+p,nr-p}=
\alpha^{r+s}\frac{pr}{r+s}
\binom{s+\frac{p}{n}}{s}\binom{r-\frac{p}{n}}{r}=q_{ns+p,nr-p}.
\non
\ena
Thus the proposition is proved.
\qed

\begin{corollary}\label{bidif-cor}
\vskip1mm
\noindent
(i)
\bea
&&
e^{q([z_1]|[z_2])}=\frac{H(z_1,z_2)}{z_2-z_1}.
\label{bidif-cor-1}
\ena
\noindent
(ii) \bea
&&
e^{q([z]|[z])}=(1-\alpha z^n})^{\frac{1-n}{n}.
\label{bidif-cor-2}
\ena
\end{corollary}
\vskip1mm
\noindent
{\it Proof.} 
(i) Integrating (\ref{bidif-1}) twice and using 
\bea
&&
\frac{H(z_1,0)}{-z_1}=\frac{H(0,z_2)}{z_2}=1,
\non
\ena
we get the result. 
\par
\noindent
(ii) is obtained by taking the limit $z_2\to z_1$ in (i).
\qed

\section{Soliton solution on the non-zero constant background}
Applying vertex operators to $\tau_0(t;V_2(z))$ it is possible to obtain soliton  
solutions on non-zero backgrounds. In this section we compute them explicitly.

Let us consider the vertex operator \cite{DJKM},
\bea
&&
X(p,q)=e^{\sum_{m=1}^\infty t_m(p^m-q^m)}e^{\sum_{m=1}^\infty(-p^{-m}+q^{-m})\frac{\partial_m}{m}},
\quad
\partial_m=\frac{\partial}{\partial t_m}.
\label{v-1}
\ena
Then
\bea
&&
X(p_1,q_1)X(p_2,q_2)=\frac{(p_1-p_2)(q_1-q_2)}{(p_1-q_2)(q_1-p_2)}:X(p_1,q_1)X(p_2,q_2):,
\label{v-2}
\ena
where the normal ordering symbol $:\quad:$ signifies to move differential operators $\partial_n$ to 
the right. It then follows the following properties,
\bea
X(p_1,q_1)X(p_2,q_2)&=&X(p_2,q_2)X(p_1,q_1) 
\quad\text{for any $p_i,q_j,$}
\label{v-3}
\\
X(p_1,q_1)X(p_2,q_2)&=&0
\quad \text{if $p_1=p_2$ or $q_1=q_2$}.
\label{v-4}
\ena

Let $M, N$ be two positive integers, $p_i, q_j, a_{i,j}$, $1\leq i\leq M$, $1\leq j\leq N$ complex parameters such that 
$\{p_i, q_j\}$ are all nonzero and mutually distinct.
Consider the operator 
\bea
&&
g=e^{\sum_{i=1}^M\sum_{j=1}^N a_{i,j} X(p_i,q_j)}.
\label{v-5}
\ena
It is well known \cite{DJKM} that, for a solution $\tau({\bf t})$ of the KP-hierarchy,
$g\tau({\bf t})$ is a solution of the KP-hierarchy if it is well defined as a formal power series 
in $t$. In particular soliton solutions are obtained by  taking  $\tau({\bf t})=1$ \cite{DJKM}. 
Here we compute $g\tau_0({\bf t};V_n(z))$.

Let us set $[M]=\{1,...,M\}$. For a nonnegative integer $l$ we denote $\binom{[M]}{l}$ the set of all $l$-element
subsets of $[M]$. If $I=\{i_1,...,i_l\}\in \binom{[M]}{l}$, we assume $i_1<\cdots<i_l$ unless otherwise stated.

Set, in general, 
\bea
&&
\Delta(x_1,...,x_N)=\prod_{i<j}(x_i-x_j),
\qquad
\eta({\bf t};z)=\sum_{i=1}^\infty t_i z^i.
\non
\ena
For $I\in \binom{[M]}{l}$ we define
\bea
&&
\eta_I(p)=\sum_{i\in I}\eta({\bf t};p_i), 
\non
\ena
and similarly for $\eta_J(q)$, $J\in \binom{[N]}{l}$.
We denote by $J^c\in \binom{[N]}{N-l}$ the complement of $J$ in $[N]$ for  $J\in \binom{[N]}{l}$.

Set 
\bea
&&
\tilde{a}_{i,j}=a_{i,j}\prod_{m=1, m\neq j}^{N}\frac{q_j-q_m}{p_i-q_m}.
\label{a-tilde}
\ena
Define the $N\times (N+M)$ matrix $B=(b_{i,j})$ by
\bea
B=
\left(
\begin{array}{ccccccc}
1&\ldots&0&\tilde{a}_{1,1}&\ldots&\tilde{a}_{M,1}\\
\vdots&\ddots&\vdots&\vdots&\quad&\vdots\\
0&\ldots&1&\tilde{a}_{N,1}&\ldots&\tilde{a}_{M,N}\\
\end{array}
\right).
\non
\ena
For $K=(k_1,...,k_N)\in \binom{[M+N]}{N}$ $B_K$ denotes the minor determinant corresponding to the 
columns specified by $K$:
\bea
&&
B_K=\det(b_{i,k_j})_{1\leq i,j\leq N}.
\non
\ena
Set
\bea
&&
(\kappa_1,...,\kappa_{M+N})=(q_1,...,q_N,p_1,...,p_M).
\non
\ena
For $K\in \binom{[M+N]}{N}$ define 
\bea
&&
\eta_K(\kappa)=\sum_{k\in K}\eta({\bf t};\kappa_k),
\quad
\eta_K(\kappa^{-1})=\sum_{k\in K}\eta({\bf t};\kappa_k^{-1}).
\non
\\
&&
\Delta_K(\kappa)=\prod_{i<j,i,j\in K}(\kappa_i-\kappa_j).
\non
\ena

To make the result simple we introduce the function 
\bea
\tau^{(n)}(g|{\bf t})=\Delta(q_1,...,q_N)e^{\sum_{j=1}^N \eta({\bf t};q_j)}g\tau_0({\bf t}-\sum_{j=1}^{N}[q_j^{-1}];V_n(z)),
\label{v-6}
\ena
which is obviously a solution of the KP-hierarchy.

Then

\begin{theorem}\label{vop-main} 
Let $p_i, q_j, a_{i,j}$ are complex parameters such that $\{p_i,q_j\}$ are all non-zero and mutually distinct.
Then 
\bea
&&
\tau^{(n)}(g|{\bf t})=\tilde{\tau}^{(n)}(g|{\bf t})\tau_0({\bf t};V_n(z)).
\label{v-7}
\ena
Here
\bea
&&
\tilde{\tau}^{(n)}(g|{\bf t})
=
\sum_{I\in \binom{[M+N]}{N}} B_I C_I(\kappa) 
e^{\eta_I(\kappa)-\sum_{i\in I}q({\bf t}|[\kappa_i^{-1}])},
\label{v-8}
\\
&&
C_I(\kappa)
=
\prod_{i<j,i,j\in I}\left(\kappa_i\kappa_j
H(\kappa_i^{-1},\kappa_j^{-1})
\right)
\prod_{i\in I}(1-\alpha \kappa_i^{-n})^{\frac{1-n}{n}}.
\non
\ena
\end{theorem}
\vskip1mm
\noindent
{\it Proof.} Using (\ref{v-2})-(\ref{v-4}) we have
\bea
g
&=&
\prod_{j=1}^N\left(1+\sum_{i=1}^M a_{i,j}X(p_i,q_j)\right)
\non
\\
&=&
\sum_{l=0}^N\sum_{1\leq j_1<\cdots<j_l\leq N}\sum_{i_1,...,i_l=1}^M
a_{i_1,j_1}\cdots a_{i_l,j_l}
\prod_{r<s}^l
\frac{(p_{i_r}-p_{i_s})(q_{j_r}-q_{j_s})}{(p_{i_r}-q_{j_s})(q_{j_r}-p_{i_s})}
\non
\\
&&
\times
:X(p_{i_1},q_{j_1})\cdots X(p_{i_l},q_{j_l}):.
\non
\ena
Rewriting $a_{i,j}$ in terms of  $\tilde{a}_{i,j}$ using  (\ref{a-tilde}) and noting that 
the normal ordering parts are symmetric in $\{p_{i_r}\}$ and $\{q_{j_s}\}$ respectively
 we get
\bea
&&g=
\sum_{l=0}^N\sum_{J=\{j_1,...,j_l\}\in \binom{[N]}{l}}\sum_{I=\{i_1,...,i_l\}\in \binom{[M]}{l}}
B_{(J^c,I)}
\frac{\Delta(q_{j_1'},...,q_{j_{N-l}'},p_{i_1},...,p_{i_l})}{\Delta(q_1,...,q_N)}
\non
\\
&&
\times
:X(p_{i_1},q_{j_1})\cdots X(p_{i_l},q_{j_l}):,
\non
\ena
where $J^c=\{j_1',...,j_{N-l}'\}\in\binom{[N]}{N-l}$ and $(J^c,I)=(j_1',...,j_{N-l}',i_1,...,i_l)$.
On the other hand 
\bea
&&
e^{\sum_{j=1}^N \eta({\bf t};q_j)}
:X(p_{i_1},q_{j_1})\cdots X(p_{i_l},q_{j_l}):\tau_0({\bf t}-\sum_{j=1}^N[q_j^{-1}];V_n(z))
\non
\\
&=&
e^{\eta_I(p)+\eta_{J^c}(q)}
\tau_0({\bf t}-\sum_{i\in I}[p_i^{-1}]-\sum_{j\in J^c}[q_j^{-1}];V_n(z)).
\non
\ena
Then
\bea
&&
\tilde{\tau}^{(n)}(g|{\bf t})
=
\sum_{l=0}^N\sum_{J=(j_1,...,j_l)\in \binom{[N]}{l}}\sum_{I=(i_1,...,i_l)\in \binom{[M]}{l}}
B_{(J^c,I)}
\Delta(q_{J^c},p_I)
\non
\\
&&\times
e^{\eta_I(p)+\eta_{J^c}(q)}
\tau_0({\bf t}-\sum_{i\in I}[p_i^{-1}]-\sum_{j\in J^c}[q_j^{-1}];V_n(z)),
\non
\ena
where $q_{J^c}=(q_{j_1'},...,q_{j_{N-l}'})$, $p_I=(p_{i_1},...,p_{i_l})$.
By considering the summation in $l,I,J$ as that over the indices $(J^c,I)$ we have
\bea
&&
\tilde{\tau}^{(n)}(g|{\bf t})
=
\sum_{K\in \binom{[M+N]}{N}}
B_{K}
\Delta_K(\kappa)
e^{\eta_K(\kappa)}
\tau_0({\bf t}-\sum_{k\in K}[\kappa_k^{-1}];V_n(z)),
\non
\ena

Now the theorem follows from

\begin{lemma}\label{vop-lemma}
For parameters $z_i$, $1\leq i\leq N$, the following equation is valid:
\bea
&&
\tau_0({\bf t}-\sum_{j=1}^N[z_j];V_n(z))
=
e^{-\sum_{j=1}^Nq({\bf t}|[z_j])}
\prod_{j=1}^N(1-\alpha z_j)^{\frac{1-n}{n}}
\prod_{i<j}^N\frac{H(z_i,z_j)}{z_j-z_i}
\tau({\bf t};V_n(z)).
\non
\ena
\end{lemma}
\vskip1mm
\noindent
{\it Proof.}
Using the bilinearity of $q({\bf t}|S)$ and the relation (\ref{tau-3}) we have
\bea
&&
\tau_0({\bf t}-\sum_{j=1}^N[z_j];V_n(z))
=e^{-\sum_{j=1}^Nq({\bf t}|[z_j])+\sum_{i\leq j}q([z_i]|[z_j])}\tau_0({\bf t};V_n(z))
\non
\ena
Then the assertion follows from Corollary \ref{bidif-cor}.
\qed

Let 
\bea
&&
u({\bf t})=2\frac{\partial^2}{\partial x^2}\log\tau^{(n)}(g|{\bf t})=
2q_{1,1}+2\frac{\partial^2}{\partial x^2}\log\tilde{\tau}^{(n)}(g|{\bf t})
\label{v-9}
\ena
be a solution of (\ref{KP-3}). By (\ref{main-1}) in Theorem \ref{main} we have
\bea
&&
q_{1,1}=\left\{
\begin{array}{ll}
\frac{\alpha}{2}&n=2\\
0&n>2.
\end{array}
\right.
\ena
Thus we have

\begin{corollary}\label{vop-cor}
Let $\tilde{\tau}^{(2)}(g|{\bf t})$ be given by (\ref{v-8}) with $n=2$. Then 
\bea
&&
u({\bf t})=\alpha+2\frac{\partial^2}{\partial x^2}\log\tilde{\tau}^{(2)}(g|{\bf t}),
\label{v-10}
\ena
is a solution of the KP equation (\ref{KP-3}).
\end{corollary}

\begin{remark} If $\alpha=0$ then $u(t)$ given by (\ref{v-10}) becomes 
a well known soliton solution \cite{DJKM,Mum1983,Kodama2017,Kodama2018,NMPZ}.
If parameters are chosen such that $\tilde{\tau}^{(2)}(g|t)$ is positive for real varaibales $t$, the second term of the right hand side of (\ref{v-10}) decays exponentially as $(t_1,t_2)$ goes to infinity except finite number of  directions. Thus  $u(t)$ can be considered as a soliton solution on a non-zero constant 
background if $\alpha\neq 0$.
\end{remark}


\begin{remark} 
For real $\kappa_j$, $1\leq j\leq M+N$, $\alpha$, $t$, the function  $\tilde{\tau}^{(n)}(g|t)$ is non-negative 
 up to overall constant if, for example,  $\kappa_1>\cdots>\kappa_{M+N}>\alpha^{1/n}\geq 0$ and $B_I\geq 0$ for all 
 $I\in \binom{[M+N]}{N}$.
\end{remark}
Here are some examples of $\tilde{\tau}^{(2)}(g|t)$.

\begin{example} In the case $N=M=1$ and $B=(1,b)$ we have
\bea
&&
\tilde{\tau}^{(2)}(g|{\bf t})
\non
\\
&&=
(1-\alpha \kappa_1^{-2})^{-\frac{1}{2}}e^{\eta({\bf t};\kappa_1)-q({\bf t}|[\kappa_1^{-1}])}
+b(1-\alpha \kappa_2^{-2})^{-\frac{1}{2}}e^{\eta({\bf t};\kappa_2)-q({\bf t}|[\kappa_2^{-1}])}.
\label{12-soliton}
\ena
\end{example}

\begin{example} In the case $N=1$, $M\geq 1$, $B=(b_1,..,b_{M+1})$, $b_1=1$, we have
\bea
&&
\tilde{\tau}^{(2)}(g|{\bf t})
=
\sum_{j=1}^{M+1} b_j (1-\alpha \kappa_j^{-2})^{-\frac{1}{2}}e^{\eta({\bf t};\kappa_j)-q({\bf t}|[\kappa_j^{-1}])}.
\label{1M-soliton}
\ena
\end{example}

\begin{example} In the case $N=M=2$ and 
\bea
&&
B=\left(
\begin{array}{cccc}
1&0&a&b\\
0&1&c&d\\
\end{array}
\right),
\non
\ena
we have
\bea
\tilde{\tau}^{(2)}(g|{\bf t})
&=&C_{12}(\kappa)e^{\eta({\bf t};\kappa_1)+\eta({\bf t};\kappa_2)-q({\bf t}|[\kappa_1^{-1}])-q({\bf t}|[\kappa_2^{-1}])}
\non
\\
&&
+cC_{13}(\kappa)e^{\eta({\bf t};\kappa_1)+\eta({\bf t};\kappa_3)-q({\bf t}|[\kappa_1^{-1}])-q({\bf t}|[\kappa_3^{-1}])}
\non
\\
&&
+dC_{14}(\kappa)e^{\eta({\bf t};\kappa_1)+\eta({\bf t};\kappa_4)-q({\bf t}|[\kappa_1^{-1}])-q({\bf t}|[\kappa_4^{-1}])}
\non
\\
&&
-aC_{23}(\kappa)e^{\eta({\bf t};\kappa_2)+\eta({\bf t};\kappa_3)-q({\bf t}|[\kappa_2^{-1}])-q({\bf t}|[\kappa_3^{-1}])}
\non
\\
&&
-bC_{24}(\kappa)e^{\eta({\bf t};\kappa_2)+\eta({\bf t};\kappa_4)-q({\bf t}|[\kappa_2^{-1}])-q({\bf t}|[\kappa_4^{-1}])}
\non
\\
&&
+(ad-bc)C_{34}(\kappa)e^{\eta({\bf t};\kappa_3)+\eta({\bf t};\kappa_4)-q({\bf t}|[\kappa_3^{-1}])-q({\bf t}|[\kappa_4^{-1}])},
\non
\ena
where
\bea
&&
C_{ij}(\kappa)=\kappa_i\kappa_jH(\kappa_i^{-1},\kappa_j^{-1})
(1-\alpha\kappa_i^{-2})^{-\frac{1}{2}}(1-\alpha\kappa_j^{-2})^{-\frac{1}{2}}.
\non
\ena
\end{example}

To study the solution as a function of $(x,y,t):=(t_1,t_2,t_3)$ the following lemma, 
which is calculated using  Corollary \ref{bidif-cor}, is useful.

\begin{lemma} For $n=2$ the following equations holds.
\vskip2mm
\noindent
(i) $\displaystyle \sum_{j=1}^\infty q_{1,j}\frac{z^j}{j}=
z^{-1}
\left(1-(1-\alpha z^2)^{\frac{1}{2}}\right)$.
\vskip2mm
\noindent
(ii) $\displaystyle \sum_{j=1}^\infty q_{2,j}\frac{z^j}{j}=0$.
\vskip2mm
\noindent
(iii) $\displaystyle \sum_{j=1}^\infty q_{3,j}\frac{z^j}{j}=
z^{-3}\left(
1-(1-\alpha z^2)^{\frac{1}{2}}(1+\frac{1}{2}\alpha z^2)
\right)$

\end{lemma}


\vskip10mm
\noindent
{\bf \large Acknowledegements}
\vskip2mm
\noindent
I would like to thank Saburo Kakei and Yasuhiro Ohta for valuable comments which they gave me
at the conference ``Varieties of  Studies on Non-Linear Waves'' held at Kyushu University
in November, 2019.  I would also like to thank Yasuhiko Yamada for useful comments on the manuscript,
and Kanehisa Takasaki for pointing out the reference \cite{Givental2003} and invaluable comments 
related with the dispertionless KP-hierarchy.

This work was supported by JSPS KAKENHI Grant Number JP19K03528.

\end{document}